\documentstyle[12pt]{article}
\input style

\catcode`@=12
\input math_macros
\begin{document}
\def\tilde{\widetilde}
\def\thefootnote{\fnsymbol{footnote}}
\FERMILABPub{93/316--T}
\begin{titlepage}
\begin{flushright}
        FERMILAB--PUB--93/316--T\\
        OSU Preprint 282\\
        October 1993\\
\end{flushright}
%\vspace{-0.20in}
\begin{center}
{\large \bf New Approach for the Construction of Fermion Mass Matrices\\
        in SO(10) SUSY GUTS}\\
%\vfill
\vskip 0.20in
        {\bf Carl H. ALBRIGHT}\\
 Department of Physics, Northern Illinois University, DeKalb, Illinois
60115\footnote{Permanent address}\\[-0.2cm]
        and\\[-0.2cm]
 Fermi National Accelerator Laboratory, P.O. Box 500, Batavia, Illinois
60510\footnote{Electronic address: ALBRIGHT@FNALV}\\
        and\\
        {\bf Satyanarayan NANDI}\\
 Department of Physics, Oklahoma State University, Stillwater, Oklahoma
        74078\footnote{Electronic address: PHYSSNA@OSUCC}\\
\end{center}
%\vskip 0.5in
\vfill
\begin{abstract}
A new procedure is developed which enables one to start from the quark and
lepton mass and mixing data at the low scale and construct mass matrices
which exhibit simple SO(10) structure at the SUSY GUT scale.  This approach
is applied to the present data involving quark and charged lepton masses, the
CKM mixing matrix and the MSW solar neutrino and
atmospheric neutrino depletion effects.  In terms of just 12 model parameters
suggested by the procedure for the 5 mass matrices, we can reproduce 15 masses
and 8 mixing parameters remarkably consistent with the input starting values.
\end{abstract}
\end{titlepage}
The apparent success$^1$ of the minimal supersymmetric standard model to
predict
unification of the gauge couplings at a scale of $10^{16}$ GeV has also spurred
renewed interest in supersymmetric grand unified models of quark and
lepton mass matrices based on SO(10).
Various model constructions have been proposed$^2$ and explored in detail$^3$
which exhibit a maximum number of
texture zeros (zeros in the upper triangular parts of the up and down quark
mass matrices),
since they may arise naturally from discrete symmetries present at the SUSY
GUT scale.  Other models have been devised$^4$ based on a minimal number of
SO(10) Higgs representations, such as one ${\bf 10}$ and one ${\bf 126}$.
Still others$^5$ propose that additional SO(10) representations be present due
to higher-dimensional operators in the Yukawa Lagrangian.

In this Letter, we propose
yet another approach which involves the construction of mass matrices from
all the known or presumed-known low-energy data.  Our procedure has the
desired feature that as more of the low-energy mass and mixing data becomes
better pinned down, the construction of the mass matrices can be repeated
and the symmetry-breaking character of the grand unified theory more clearly
illuminated.  For purposes of illustration, we apply the general method to
masses and mixings consistent with the observed MSW$^6$ solar neutrino and
atmospheric neutrino depletion effects.$^{7-8}$  The model constructed has a
number of interesting features and is able to reproduce 15 mass and 8 mixing
parameters with just 12 model input parameters.

The new approach proposed for the construction of quark and lepton mass
matrices consists of the following steps:
\begin{itemize}
\item   Start from the known and/or presumed-known quark and lepton masses,
        $m_q$'s, $m_{\ell}$'s and $m_{\nu}$'s; and quark and lepton mixing
        matrices, $V_{CKM}$ and $V_{LEPT}$, at the low scales.
\item   Evolve the masses and mixing matrices to the SUSY GUT scale
        using the appropriate renormalization group equations
        (RGEs) for the minimal supersymmetric standard model (MSSM).
\item   Construct complex symmetric $M^U$, $M^D$, $M^E$, and $M^{N_{eff}}$
        matrices for the up and down quarks, charged leptons and
        light neutrinos using a modified procedure of Kusenko$^9$ described
        later.  Two parameters $x_q$ and $x_{\ell}$ allow one to adjust the
        diagonal-- off-diagonal nature of the quark and lepton mass matrices.
\item   Vary $x_q$ and $x_{\ell}$ systematically over their support regions
        while searching for as many pure ${\bf 10}$ or pure ${\bf 126}$
        $SO(10)$ contributions to the matrix elements as possible.
\item   For the ``best'' choice of $x_q$ and $x_{\ell}$, construct a simple
        model of the mass matrices with as many texture zeros as possible.
\item   Evolve the mass eigenvalues and mixing matrices determined from the
        model at the SUSY GUT scale to the low scale and compare the results
        with the starting input data.
\end{itemize}

In order to illustrate this procedure with a concrete example, we shall apply
it to the reasonably well-known quark data$^{10}$ and assume that the lepton
data is best described by the non-adiabatic MSW resonant oscillation
depletion$^{6-7}$ of the solar electron-neutrino flux and by the depletion of
the atmospheric muon-neutrino flux$^8$ through oscillation into tau-neutrinos.
To our knowledge, no SO(10) seesaw model has been constructed which explains
both observations with these interpretations.

With $m_t^{phys} \sim 160$ GeV, we adopt as starting input the following quark
masses$^{10,11}$
$$\begin{array}{rlrl}
        m_u(1 {\rm GeV})&= 5.1\ {\rm MeV},& \qquad m_d(1 {\rm GeV})&= 8.9\
                {\rm MeV}\nonumber\\
        m_c(m_c)&= 1.27\ {\rm GeV},& \qquad m_s(1 {\rm GeV})&= 175\ {\rm MeV}
                \cr
        m_t(m_t)&= 150\ {\rm GeV},& \qquad m_b(m_b)&\simeq 4.25\ {\rm GeV}\cr
  \end{array}\eqno(1a)$$
which are evaluated at the 1 GeV scale for the light quarks and at the running
scale for the heavy quarks, together with the CKM mixing matrix$^{10}$ at the
weak scale
$$V_{CKM} = \left(\matrix{0.9753 & 0.2210 & (-0.283 -0.126i)\times 10^{-2}\cr
                -0.2206 & 0.9744 & 0.0430\cr
                0.0112 -0.0012i & -0.0412 -0.0003i & 0.9991\cr}\right)
        \eqno(1b)$$
For the leptons, we use the central values in the non-adiabatic MSW solar
neutrino conversion region:$^7$ $\delta m_{12}^2 \sim 5 \times 10^{-6}\
{\rm eV^2}$, $\sin^2 2\theta_{12} \sim 8 \times 10^{-3}$; as well as the
central values singled out in the muon-neutrino atmospheric depletion
region:$^8$ $\delta m_{23}^2 \sim 2 \times 10^{-2}\ {\rm eV^2}$,
$\sin^2 2\theta_{23}~\sim 0.5$.  Here we are assuming the ``conventional''
interpretation that solar
electron-neutrinos undergo resonant conversion into muon-neutrinos in the
sun, while muon-neutrinos oscillate into tau-neutrinos in traveling through
the atmosphere.  We then take for the lepton input
$$\begin{array}{rlrl}
        m_{\nu_e}&= 0.5 \times 10^{-6}\ {\rm eV},& \qquad m_e&= 0.511\ {\rm
                MeV}\nonumber\\
        m_{\nu_{\mu}}&= 0.224 \times 10^{-2}\ {\rm eV},& \qquad m_{\mu}&=
                105.3\ {\rm MeV}\cr
        m_{\nu_{\tau}}&= 0.141\ {\rm eV},& \qquad m_{\tau}&= 1.777\ {\rm
                GeV}\cr \end{array}\eqno(2a)$$
and
$$V_{LEPT} = \left(\matrix{0.9990 & 0.0447 & (-0.690 -0.310i) \times 10^{-2}\cr
                -0.0381 -0.0010i & 0.9233 & 0.3821\cr
                0.0223 -0.0030i & -0.3814 & 0.9241\cr}\right) \eqno(2b)$$
We have simply assumed a value for the electron-neutrino mass and constructed
the lepton mixing matrix$^{12}$ by making use of the unitarity
conditions.

We now evolve the low energy data to the SUSY GUT scale, $\Lambda_{SGUT} = 1.2
\times 10^{16}$ GeV, using numbers taken from the work of Naculich.$^{13}$  For
this
purpose and in most cases the one-loop RGEs will suffice, so one can use
analytic expressions for the running variables.  We adjust $m_b$ and $\tan
\beta = v_u/v_d$, the ratio of the up quark to the down quark VEVs, so that
complete Yukawa unification is achieved at $\Lambda_{SGUT}$, i.e.,
$\bar{m}_{\tau} = \bar{m}_b = \bar{m}_t/\tan \beta$.  This is accomplished by
choosing $m_b(m_b) = 4.09$ GeV at the running $b$ quark mass scale$^{14}$ and
$\tan \beta = 48.9$.  The evolved masses at $\Lambda_{SGUT}$ are then found to
be
$$\begin{array}{rlrl}
        \bar{m}_u&= 1.098\ {\rm MeV},& \qquad \bar{m}_d&= 2.127\
                {\rm MeV}\nonumber\\
        \bar{m}_c&= 0.314\ {\rm GeV},& \qquad \bar{m}_s&= 42.02\ {\rm MeV}\cr
        \bar{m}_t&= 120.3\ {\rm GeV},& \qquad \bar{m}_b&= 2.464\ {\rm GeV}\cr
        \bar{m}_{\nu_e}&= 0.581 \times 10^{-6}\ {\rm eV},& \qquad \bar{m}_e&=
                0.543\ {\rm MeV}\nonumber\\
        \bar{m}_{\nu_{\mu}}&= 0.260 \times 10^{-2}\ {\rm eV},& \qquad
                \bar{m}_{\mu}&= 111.9\ {\rm MeV}\cr
        \bar{m}_{\nu_{\tau}}&= 0.164\ {\rm eV},& \qquad \bar{m}_{\tau}&=
                2.464\ {\rm GeV}\cr \end{array}\eqno(3a)$$
The following $V_{CKM}$ and $V_{LEPT}$ mixing matrix elements also evolve to
$$\begin{array}{rlrl}
        \bar{V}_{ub}&= (-0.2163 -0.0963i) \times 10^{-2},&\qquad
                \bar{V}_{13}&= (-0.634 -0.285i) \times 10^{-2}\nonumber\\
        \bar{V}_{cb}&= 0.0329&\qquad \bar{V}_{23}&= 0.3508\cr
        \bar{V}_{td}&= 0.0086 -0.0009i&\qquad \bar{V}_{31}&= 0.0205 -0.0028i\cr
        \bar{V}_{ts}&= -0.0315 -0.0002i&\qquad \bar{V}_{32}&= -0.3502\cr
                \end{array}\eqno(3b)$$
while the other mixing matrix elements receive smaller corrections
which can be neglected.

In order to construct the quark mass matrices at the SUSY GUT scale from the
above information, we use a procedure due to Kusenko$^{9}$ modified for
our purposes.  One expresses the unitary CKM mixing matrix in terms of one
Hermitian generator by writing $V_{CKM} = U'_L U^{\dagger}_L = \exp(i\alpha
H)$, where
$$i\alpha H = \sum^3_{k=1}(\log v_k){{\prod_{i\neq k}(V_{CKM} - v_i I)}\over{
        \prod_{i\neq k}(v_k - v_i)}}$$
and the $v_j$ are the eigenvalues of the unitary mixing matrix.
The transformation matrices from the weak to the mass bases are given by
$$U'_L = \exp(i\alpha Hx_q),\qquad U_L = \exp\left[i\alpha H(x_q - 1)\right]$$
where for $x_q = 0$ the up quark mass matrix is diagonal, while for $x_q = 1$
the down quark mass matrix is diagonal.  Finally, the complex-symmetric mass
matrices are determined by
$$M^U = U'^{\dagger}_L D^U U'^{\dagger T}_L,\qquad M^D = U^{\dagger}_L D^D
        U^{\dagger T}_L$$
where $D^U$ and $D^D$ are the diagonal matrices in the mass bases with entries
taken from (3a).  It suffices
to expand $V_{CKM},\ U'_L$ and $U_L$ to third order in $\alpha$ in order to
obtain accurate expressions for the mass matrices.  The lepton mass matrices,
$M^E$ for the charged leptons and $M^{N_{eff}}$ for the light neutrinos,
are constructed in a similar fashion with the parameter $x_q$ replaced
by $x_{\ell}$.

The SO(10) Yukawa interaction Lagrangian for the non-supersymmetric fermions
is given by
$${\cal L}_Y = -\sum_i{\bar{\psi^c}}^{(16)}f^{(10_i)}\psi^{(16)}
		\phi^{(10_i)} -\sum_j{\bar{\psi^c}}^{(16)}f^{(126_j)}
		\psi^{(16)}{\bar{\phi}}^{(126_j)} + {\rm h.c.}\eqno(4a)$$
where the $f$'s represent Yukawa coupling matrices and we assume just
${\bf 10}$ and ${\bf 126}$ contributions which are symmetric.$^{15}$  The mass
matrices are given by
$$\begin{array}{rl}
        M^U&= \sum_i f^{(10_i)}v_{ui} + \sum_j f^{(126_j)}w_{uj}\nonumber\\
        M^D&= \sum_i f^{(10_i)}v_{di} + \sum_j f^{(126_j)}w_{dj}\cr
        M^{N_{Dirac}}&= \sum_i f^{(10_i)}v_{ui} - 3\sum_j f^{(126_j)}w_{uj}\cr
        M^E&= \sum_i f^{(10_i)}v_{di} - 3\sum_j f^{(126_j)}w_{dj}\cr
                \end{array}\eqno(4b)$$
where $v_{ui}$ and $w_{uj}$ are the ${\bf 10}$ and ${\bf 126}$ VEV
contributions to the up quark and Dirac neutrino matrices, and similarly
for the down quark and charged lepton contributions.  The equations in (4b)
can be inverted to determine the sum of the ${\bf 10}$ and sum of the ${\bf
126}$ contributions separately.  At this stage we do not know how many
${\bf 10}$ and ${\bf 126}$ representations of each type are necessary.

By varying the $x_q$ and $x_{\ell}$ parameters over the unit square support
region and by allowing all possible signs to appear in the diagonal matrix
entries of $D^U,\ D^D,\ D^E$ and $D^{N_{eff}}$, we can search for a set of
mass matrices which have the simplest
${\bf 10}$ - ${\bf 126}$ structure for as many matrix elements as possible.
Such a preferred choice is found with $x_q = 0$ and $x_{\ell} = 0.88$, where
the observed structure for $M^U$ is diagonal and
$$M^D \sim M^E \sim \left(\matrix{10,126 & 10,126 & 10\cr 10,126 & 126 & 10\cr
        10 & 10 & 10\cr}\right)\eqno(5a)$$
with $M^D_{11},\ M^E_{12}$ and $M^E_{21}$ anomalously small.  We shall assume
these elements, in fact, exhibit texture zeros and also assume that
the same ${\bf 10}$ and ${\bf 126}$ contribute, respectively, to the 33 and 22
diagonal elements of $M^U$ and $M^D$.  Hence
$$M^U \sim M^{N_{Dirac}} \sim diag(10,126;\ 126;\ 10)\eqno(5b)$$
If we require as simple a structure as possible for the 11 elements of the
four matrices, we are led numerically to the following choices for the Yukawa
coupling matrices at $\Lambda_{SGUT}$
$$\begin{array}{rlrl}
f^{(10)}&= diag(0,\ 0,\ f^{(10)}_{33}),&\qquad f^{(126)}&= diag(f^{(126)}_{11},
        \ f^{(126)}_{22},\ 0)\nonumber\\[0.1in]
f^{(10')}&= \left(\matrix{f^{(10')}_{11}&f^{(10')}_{12}&f^{(10')}_{13}\cr
        f^{(10')}_{12}& 0 &f^{(10')}_{23}\cr
        f^{(10')}_{13}&f^{(10')}_{23}& 0\cr}\right),&\qquad
f^{(126')}&= \left(\matrix{0 & f^{(126')}_{12}& 0\cr
        f^{(126')}_{12}& 0 & 0\cr 0 & 0 & 0\cr}\right)\end{array}\eqno(6)$$
The model requires two ${\bf 10}$'s and two ${\bf 126}$'s of SO(10) with
${\bf 10'}$ and ${\bf 126'}$ having no VEVs in the up direction.  There are
four
texture zeros in the $M^U$ and $M^D$ matrices taken together.
The four matrices assume the simple textures
$$\eqalignno{
        M^U&= f^{(10)}v_u + f^{(126)}w_u\ = diag(F',\ E',\ C')&(7a)\cr
        M^{N_{Dirac}}&= f^{(10)}v_u - 3 f^{(126)}w_u = diag(-3F',\ -3E',\ C')
                &(7b)\cr
        M^D&= f^{(10)}v_d + f^{(126)}w_d + f^{(10')}v'_d + f^{(126')}w'_d\cr
           &= \left(\matrix{0 & A & D\cr A & E & B\cr D & B & C\cr}\right)
                &(7c)\cr
        M^E&= f^{(10)}v_d - 3 f^{(126)}w_d + f^{(10')}v'_d - 3 f^{(126')}w'_d
                \cr
           &= \left(\matrix{F & 0 & D\cr 0 & -3E & B\cr D & B & C\cr}\right)
                &(7d)\cr}$$
with only $D$ complex and the following relations holding
$$\begin{array}{rlrl}
        C'/C&= v_u/v_d,&\qquad E'/E&= w_u/w_d\nonumber\\
        f^{(10')}_{11}v'_d&= -f^{(126)}_{11}w_d = {1\over{4}}F,&\qquad
        f^{(126)}_{11}w_u&= F'\cr
        f^{(10')}_{12}v'_d&= 3f^{(126')}_{11}w'_d = {3\over{4}}A&&\cr
                \end{array}\eqno(7e)$$
from which we obtain the constraint, $4F'/F = -E'/E$.

With $F' = -\bar{m}_u,\ E' = \bar{m}_c,\ C' = \bar{m}_t$
$$\begin{array}{rlrl}
        C&= 2.4607,& \qquad &{\rm so}\quad v_u/v_d = \tan \beta = 48.9
                \nonumber\\
        E&= -0.3830 \times 10^{-1},& \qquad &{\rm hence}\quad w_u/w_d =
-8.20\cr
        F& = -0.5357 \times 10^{-3},& \qquad B&= 0.8500 \times 10^{-1}\cr
        A& = -0.9700 \times 10^{-2},& \qquad D&= (0.4200 + 0.4285i) \times
                10^{-2}\cr \end{array}\eqno(8)$$
the masses and mixing matrices are calculated at $\Lambda_{SGUT}$ by use of
the projection operator technique of Jarlskog$^{16}$ and then evolved to the
low scales.  The following low-scale results emerge for the quarks:
$$\begin{array}{rlrl}
        m_u(1 {\rm GeV})&= 5.10\ {\rm MeV},& \qquad m_d(1 {\rm GeV})&= 9.33
                \ {\rm MeV}\nonumber\\
        m_c(m_c)&= 1.27\ {\rm GeV},& \qquad m_s(1 {\rm GeV})&= 181\ {\rm MeV}
                \cr
        m_t(m_t)&= 150\ {\rm GeV},& \qquad m_b(m_b)&= 4.09\ {\rm GeV}\cr
  \end{array}\eqno(9a)$$
$$V_{CKM} = \left(\matrix{0.9753 & 0.2210 & (0.2089 -0.2242i)\times 10^{-2}\cr
                -0.2209 & 0.9747 & 0.0444\cr
                0.0078 -0.0022i & -0.0438 -0.0005i & 0.9994\cr}\right)
        \eqno(9b)$$
These results are in excellent agreement with the input in (1a,b), aside from
the unknown CP phase, with$^{10,11}$ $|V_{ub}/V_{cb}| = 0.069$ and
$m_s/m_d = 19.4$.

For the leptons we observe that the heavy righthanded Majorana neutrino mass
matrix can be computed at $\Lambda_{SGUT}$ from the approximate seesaw mass
formula$^{17}$
$$M^R = - M^{N_{Dirac}}(M^{N_{eff}})^{-1}M^{N_{Dirac}}\eqno(10a)$$
and numerically can be approximated by the nearly geometric form
$$M^R = \left(\matrix{F'' & - {2\over{3}}\sqrt{F''E''} &
                -{1\over{3}}\sqrt{F''C''}e^{i\phi_{D''}}\cr
                - {2\over{3}}\sqrt{F''E''} & E'' &
                        -{2\over{3}}\sqrt{E''C''}e^{i\phi_{B''}}\cr
                -{1\over{3}}\sqrt{F''C''}e^{i\phi_{D''}} &
                -{2\over{3}}\sqrt{E''C''}e^{i\phi_{B''}} & C''\cr}\right)
                        \eqno(10b)$$
where $E'' = {2\over{3}}\sqrt{F''C''}$ and $\phi_{B''} = - \phi_{D''}/3$;
moreover, the structure in (10b) can be separated into two parts with
coefficients $2/3$ and $1/3$ which suggests they may arise again from two
different ${\bf 126}$ contributions.  Such geometric textures have been
studied at some length by Lemke.$^{18}$

With $C'' = 0.6077 \times 10^{15},\ F'' = 0.1745\times 10^{10}$ and
$\phi_{D''} = 45^o$, $M^R$ is reproduced exceedingly well at
$\Lambda_{SGUT}$ with the resulting heavy Majorana neutrino masses
$M_{R_1} = 0.249 \times 10^9$~GeV, $M_{R_2} = 0.451\times 10^{12}$ GeV and
$M_{R_3} = 0.608\times 10^{15}$ GeV.  At the low scales we find
$$\begin{array}{rlrl}
        m_{\nu_e}&= 0.534 \times 10^{-5}\ {\rm eV},& \qquad m_e&= 0.504\ {\rm
                MeV}\nonumber\\
        m_{\nu_{\mu}}&= 0.181 \times 10^{-2}\ {\rm eV},& \qquad m_{\mu}&=
                105.2\ {\rm MeV}\cr
        m_{\nu_{\tau}}&= 0.135\ {\rm eV},& \qquad m_{\tau}&= 1.777\ {\rm
                GeV}\cr \end{array}\eqno(11a)$$
and
$$V_{LEPT} = \left(\matrix{0.9990 & 0.0451 & (-0.029 -0.227i) \times 10^{-2}
		\cr -0.0422 & 0.9361 & 0.3803\cr
                0.0174 -0.0024i & -0.3799 -0.0001i & 0.9371\cr}\right)
                \eqno(11b)$$

The agreement with our starting input is remarkably good, especially since
only 12 model parameters have been introduced in order to explain 15 masses
and 8 effective mixing parameters.  Although we need two ${\bf 10}$ and
two ${\bf 126}$ Higgs representations for the up, down, charged lepton and
Dirac neutrino matrices with one or two additional ${\bf 126}$'s for the
Majorana matrix, pairs of irreducible representations more naturally emerge
in the superstring framework than do single Higgs representations.  We have
thus demonstrated by the model constructed that all quark and lepton mass and
mixing data (as assumed herein) can be well understood in the framework of a
simple SUSY GUT model based on SO(10) symmetry.

While we have gone into some detail about the model which has been constructed
based on the low energy data involving the solar neutrino and atmospheric
neutrino depletions, we wish to emphasize that the same approach can be
carried out for other starting points.  In a paper$^{19}$ to be
published elsewhere, we shall elaborate on the numerical details leading to
the solution presented here and consider alternative scenarios for the lepton
masses and mixing matrix.

The authors gratefully acknowledge the Summer Visitor Program and warm
hospitality of the Fermilab Theoretical Physics Department, which enabled the
initiation of this research.  We thank Joseph Lykken for his continued interest
and comments during the course of this work.  The research of CHA was
supported in part by Grant No. PHY-9207696 from the National Science
Foundation, while that of SN was supported in part by the U.S. Department of
Energy, Grant No. DE-FG05-85ER 40215.

\newpage
\begin{reflist}
\item   U. Amaldi, W. de Boer and H. Funstenau, Phys. Lett. B {\bf 260},
                447 (1991); J. Ellis, S. Kelley and D. V. Nanopoulos,
                Phys. Lett. B {\bf 260}, 131 (1991); P. Langacker and M. Luo,
                Phys. Rev. D {\bf 44}, 817 (1991).

\item 	J. A. Harvey, P. Ramond, and D. B. Reiss, Phys. Lett {\bf 92B}, 309
		(1980); S. Dimopoulos, L. J. Hall and S. Raby, Phys. Rev. Lett.
		{\bf 68}, 1984 (1992);  Phys. Rev. D {\bf 45}, 4192 (1992);
		{\bf 46}, R4793 (1992); {\bf 47}, R3702 (1993);  G. F. Giudice,
		Mod. Phys. Lett. A {\bf 7}, 2429 (1992); H. Arason, D. J.
		Casta\~{n}o, E.-J. Piard and P. Ramond, Phys. Rev. D {\bf 47},
		232 (1993); P. Ramond, R. G. Roberts and G. G. Ross, Nucl.
		Phys. {\bf B406}, 19 (1993).

\item   V. Barger, M. S. Berger, T. Han and M. Zralek, Phys. Rev.
                Lett. {\bf 68}, 3394 (1992); Phys. Rev. D {\bf 47}, 2038
(1993).

\item   K. S. Babu and R. N. Mohapatra, Phys. Rev. Lett. {\bf 70},
                2845 (1993); L. Lavoura, Carnegie-Mellon University preprint
                93-0473, June 1993.

\item   G. W. Anderson, S. Raby, S. Dimopoulos, L. J. Hall and G. D. Starkman,
                Lawrence Berkeley Laboratory preprint LBL-33531, August 1993.

\item   S. P. Mikheyev and A. Yu Smirnov, Yad Fiz. {\bf 42}, 1441 (1985)
                [Sov. J. Nucl. Phys. {\bf 42}, 913 (1986)]; Zh. Eksp. Teor.
                Fiz. {\bf 91}, 7 (1986) [Sov. Phys. JETP {\bf 64}, 4
                (1986)]; Nuovo Cimento {\bf 9C}, 17 (1986); L. Wolfenstein,
                Phys. Rev. D {\bf 17}, 2369 (1978); {\bf 20}, 2634 (1979).

\item	R. Davis et al., Phys. Rev. Lett. {\bf 20}, 1205 (1968); in {\it
	Neutrino '88}, ed. J. Schnepp et al. (World Scientific, 1988);
	K. Hirata et al., Phys. Rev. Lett. {\bf 65}, 1297, 1301 (1990);
	A. I. Abazov et al., Phys. Rev. Lett. {\bf 67}, 3332 (1991);
	Phys. Rev. C {\bf 45}, 2450 (1992); P. Anselmann et al., Phys. Lett.
	B {\bf 285}, 376, 390 (1992).

\item   K. S. Hirata et al., Phys. Lett. B {\bf 205},416 (1988); D. Casper
                et al., Phys. Rev. Lett. {\bf 66}, 2561 (1991);

\item   A. Kusenko, Phys. Lett. B {\bf 284}, 390 (1992).

\item   Review of Particle Properties, Phys. Rev. D {\bf 45}, No. 11,
                Part II, June (1992).

\item   J. Gasser and H. Leutwyler, Phys. Rep. C {\bf 87}, 77 (1982).

\item   C. H. Albright, Phys. Rev. D {\bf 45}, R725 (1992); Zeit. f. Phys.
                C {\bf 56}, 577 (1992); A. Yu. Smirnov, Institute for Advanced
		Study and ICTP preprint, IASSNS-AST-93-14.

\item   S. Naculich, Johns Hopkins University preprint JHU-TIPAC-930002,
                January 1993.

\item   The two loop RGEs would lead to somewhat different
                $\tan \beta$ and $m_b(m_b)$.

\item   We neglect possible ${\bf 120}$ Higgs contributions which would give
                rise to antisymmetric mass matrices.

\item   C. Jarlskog, Phys. Rev. D {\bf 35}, 1685 (1987); {\bf 36}, 2138
                (1987); C. Jarlskog and A. Kleppe, Nucl. Phys. {\bf B286},
                245 (1987).

\item   M. Gell-Mann, P. Ramond, and R. Slansky, in {\it Supersymmetry},
                edited by P. Van Nieuwenhuizen and D. Z. Freedman
                (North-Holland, Amsterdam, 1979); T. Yanagida, Prog. Theor.
                Phys. {\bf B 315}, 66 (1978).

\item   E. H. Lemke, Mod. Phys. Lett. A {\bf 7}, 1175 (1992).

\item   C. H. Albright and S. Nandi, in preparation.
\end{reflist}
\end{document}